# Development of a CsI:Tl calorimeter subsystem for the All-Sky Medium-Energy Gamma-Ray Observatory (AMEGO)

Richard S. Woolf, *Member, IEEE*, J. Eric Grove, Bernard F. Phlips, and Eric A. Wulf

***Abstract*–We report on the development of the thallium-doped cesium iodide (CsI:Tl) calorimeter subsystem for the All-Sky Medium-Energy Gamma-ray Observatory (AMEGO) [1]. The CsI:Tl calorimeter is one of the three main subsystems that comprise the AMEGO instrument suite; the others include the double-sided silicon strip detector (DSSD) tracker/converter and a cadmium zinc telluride (CZT) calorimeter. Similar to the LAT instrument on Fermi Gamma-ray Space Telescope, the hodoscopic calorimeter consists of orthogonally layered CsI:Tl bars. Unlike the LAT, which uses PIN photodiodes, the scintillation light from each end of the CsI:Tl bar is read out with recently developed large-area silicon photomultiplier (SiPM) arrays. Development of the calorimeter technology for a large space-based γ-ray observatory is being done via funding from the NASA APRA program. Under this program, we have designed, built and are currently testing a prototype consisting of 24 CsI:Tl bars (each with dimensions of 16.7 mm x 16.7 mm x 100 mm) hodoscopically arranged in four layers with six bars per layer. The ends of each bar are read out with SensL ArrayJ quad SiPMs. Signal readout and processing is done with the IDEAS SIPHRA (IDE3380) ASIC. Performance testing of this prototype was done with laboratory sources and at an initial beam test; future testing and characterization will be done during a second beam test (2019) and a balloon flight (2020) in conjunction with the other subsystems led by NASA-GSFC. We will discuss the initial calibration tests, construction of the prototypes and the most up-to-date results from laboratory and beam tests with analog/digital electronics and the SIPHRA DAQ.***

## I. Introduction

With funding from the NASA Astrophysics and Analysis (APRA) program, we have built and tested a hodoscopic CsI:Tl scintillating-crystal calorimeter for a medium-energy gamma-ray Compton and pair telescope. The design and technical approach for this calorimeter relies deeply on heritage from the Fermi Large Area Telescope (LAT) CsI:Tl Calorimeter [2], but it dramatically improves the low-energy performance of that design by reading out the scintillation light with silicon photomultipliers (SiPMs) [3], making the technology developed for Fermi applicable in the Compton regime.

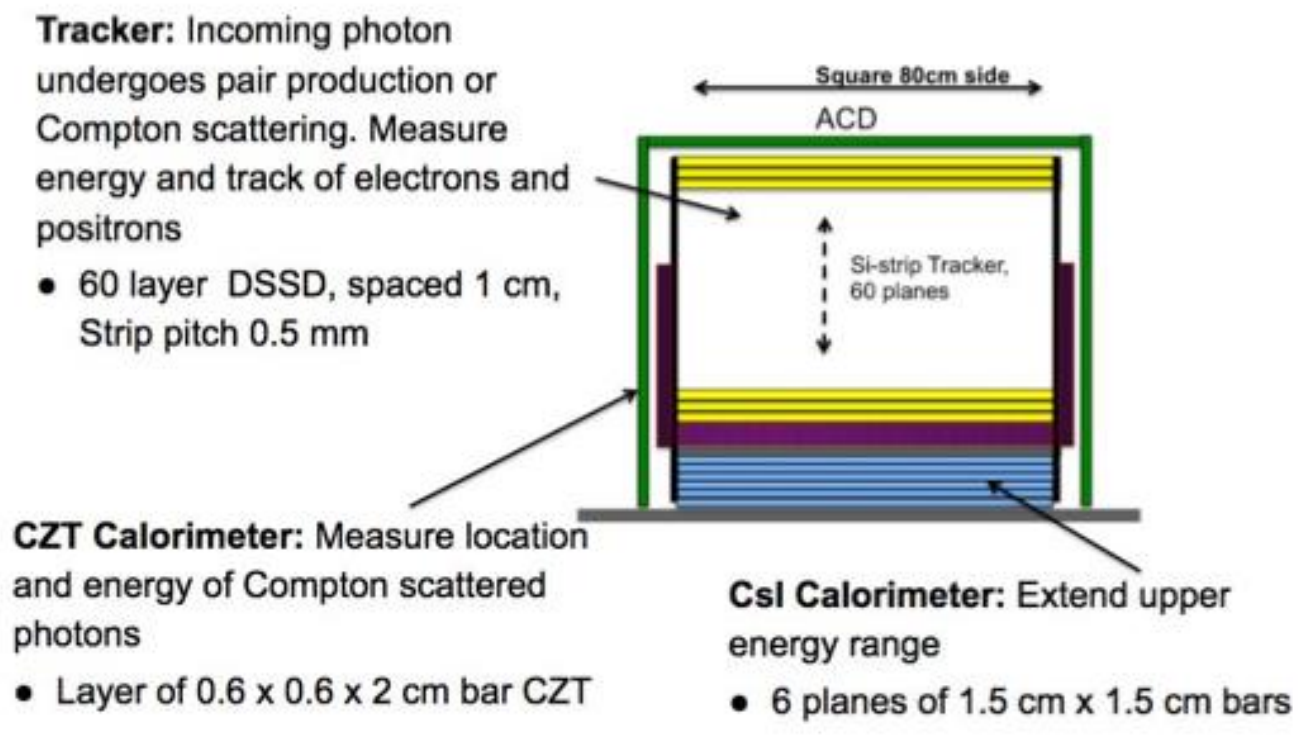

Fig. 1. Sketch of the AMEGO Probe-class mission concept. The focus of this paper is the CsI:Tl calorimeter, shown in blue.

The All-sky Medium-Energy Gamma-ray Observatory (AMEGO) is a Probe-class mission concept to perform a high-sensitivity survey of the γ-ray sky from ~0.3 MeV to ~10 GeV. AMEGO was designed to provide a dramatic increase in sensitivity relative to previous instruments in this energy range (predominantly INTEGRAL/SPI [4] and CGRO COMPTEL [5]), with the same transformative sensitivity increase – and corresponding scientific return – that the Fermi LAT [6] provided relative to CGRO EGRET [7]. To enable transformative science over a broad range of MeV energies and with a wide field of view, AMEGO is a combined Compton telescope and pair telescope employing a silicon-strip tracker (for Compton scattering and pair conversion and tracking) and a solid-state CdZnTe calorimeter (for Compton absorption) and CsI calorimeter (for pair calorimetry), surrounded by a plastic scintillator anti-coincidence detector (ACD).

## II. All-Sky Medium-Energy Gamma-Ray Observatory

### A. Instrument Concept

The AMEGO concept is illustrated in Fig. 1. As shown, the observatory consists of three major subsystems: double-sided silicon strip detectors (DSSD) which comprise the tracker (yellow); the CZT calorimeter (purple) directly beneath and on the sides of the tracker; and a deep CsI:Tl calorimeter beneath the DSSD tracker and CZT calorimeter (blue). The whole instrument is surrounded on all sides, except on the spacecraft bus side, with an ACD (green) to veto charged particles.

Manuscript received December 15, 2018. This work was sponsored by NASA-APRA (NNH14ZDA001N-APRA and NNH15ZDA001N-APRA).
R. S. Woolf is with the U. S. Naval Research Laboratory, Washington, DC 20375 USA (telephone: 202-404-2886, e-mail: richard.woolf@nrl.navy.mil).
J. E. Grove is with the U. S. Naval Research Laboratory, Washington, DC 20375 USA (telephone: 202-767-3112, e-mail: eric.grove@nrl.navy.mil).
B. F. Phlips is with the U. S. Naval Research Laboratory, Washington, DC 20375 USA (telephone: 202-767-3572, e-mail: phlips@nrl.navy.mil).
E. A. Wulf is with the U. S. Naval Research Laboratory, Washington, DC 20375 USA (telephone: 202-404-1475, e-mail: wulf@nrl.navy.mil).

We determined the position resolution (σ) and energy resolution (% equiv. Gaussian σ) as a function of energy based on measurements using laboratory sources and cosmic-ray muons with laboratory electronics. The interaction location ($z$) is calculated $z = [(R-L)/(R+L)]$, where $L$ and $R$ are the gain-corrected, pedestal-subtracted pulse heights from each SiPM.

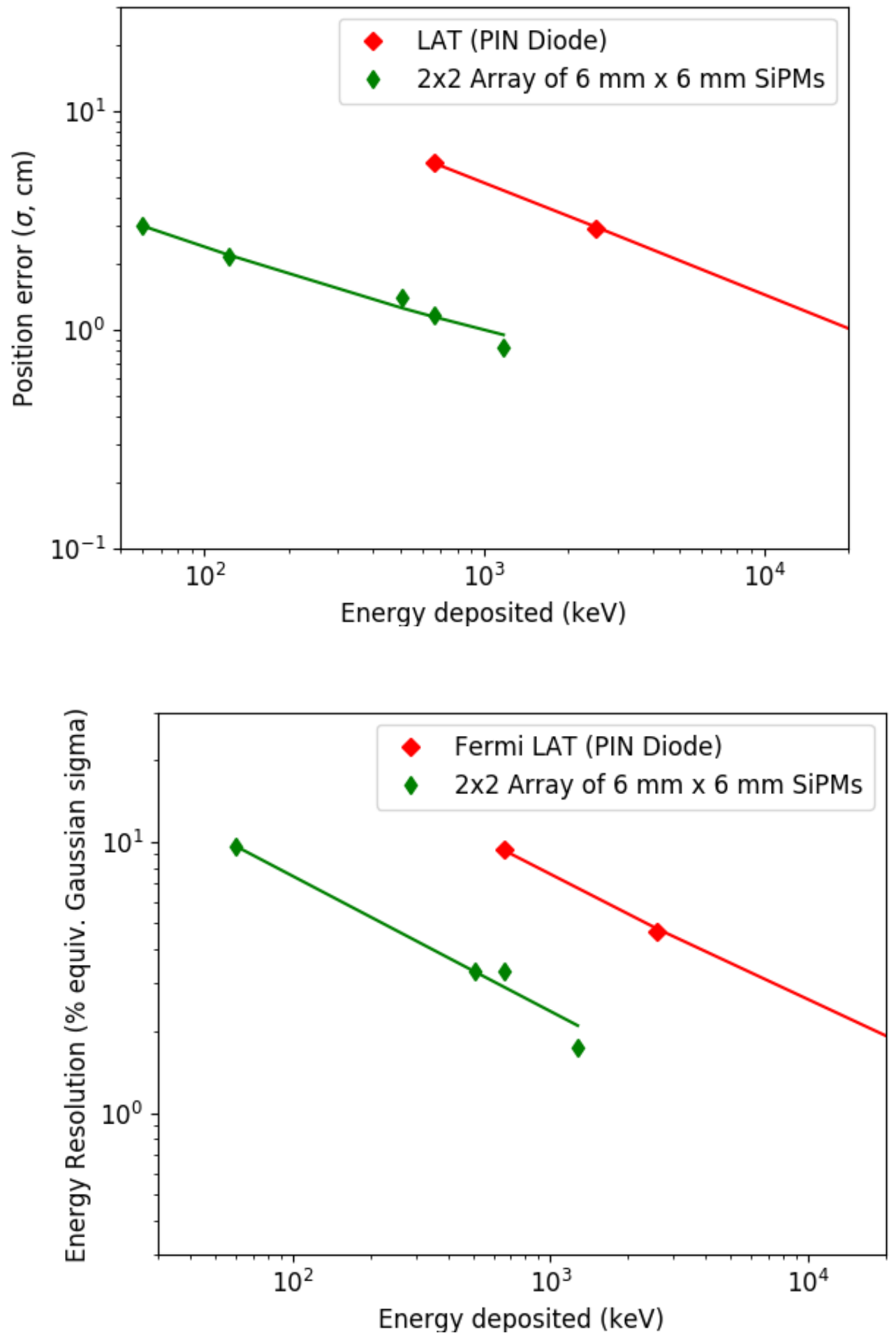

Fig. 3. Comparison of the position resolution (*top*) and energy resolution (*bottom*) as a function of energy for a Fermi LAT CDE (red) compared to a 15 mm x 15 mm x 320 mm CsI:Tl bar with SiPM readout (green). SiPM readout improves both the position (1 cm compared to 6 cm with PIN diode) and energy resolution (3% compared to 10% with PIN diode) at 662 keV.

### B. *Science Reach*

AMEGO will survey the medium-energy sky in γ rays, with particular emphasis on continuum sensitivity, spectroscopy, and polarization. The main science objects that AMEGO will concentrate on are: (1) understanding the formation, evolution, and physics of astrophysical jets, (2) identifying the physical processes in the extreme conditions around compact objects, (3) testing predictions of Dark Matter, and, (4) studying the life cycle of matter in the nearby universe.

### C. *Calorimeter Concept*

The focus of this manuscript is the CsI:Tl calorimeter subsystem, which is under development at the U.S. Naval Research Laboratory (NRL). Discussion of the other subsystems can be found elsewhere [8, 9]. The calorimeter will leverage experience from Fermi LAT calorimeter design, development, and assembly that took place at NRL in the late 1990s/early 2000s. Like the Fermi LAT calorimeter, it will be modular, comprising four identical units, with CsI:Tl crystals supported in a hodoscopic array by a mechanical structure with alternating ($x$,$y$) layers. Each crystal will be read out on each end with the newly-developed silicon photomultipliers (SiPMs) [3] as opposed to PIN diodes which were used on the Fermi LAT crystals (Fig. 2).

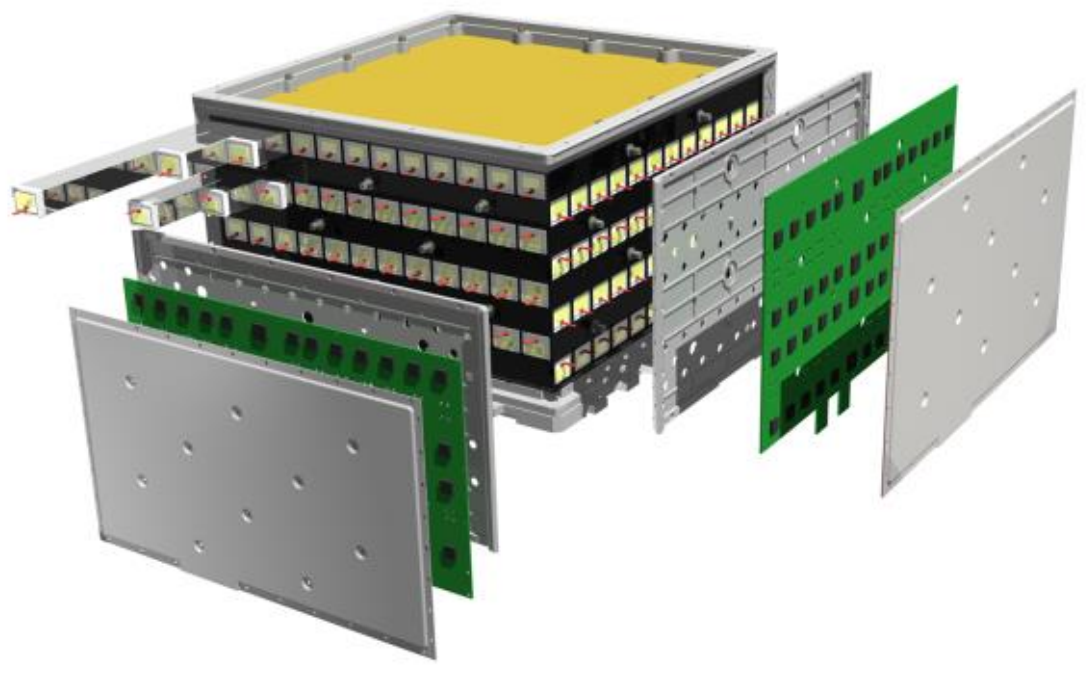

Fig. 2. Fermi LAT Calorimeter module, with CsI:Tl crystals read out at both ends and each contained in reflective wrap, in composite structure, with front-end electronics readout boards in EMI enclosures beside the array.

## III. Instrument Development

### A. *Initial Development*

Initial testing was performed with a single 15 mm x 15 mm x 320 mm CsI:Tl scintillator bar (spare from Fermi LAT prototype trade studies) in late 2015/early 2016. The bar was wrapped in Tetratex (diffusive white material) and sealed with adhesive aluminized Mylar to provide good light collection. For the optical readout, we used the SensL ArrayJ 6 mm x 6 mm quad (2x2 array) SiPMs. SiPMs are advantageous for space-based systems because they operate at low voltage (~10s of volts), provide gain comparable to a PMT (~$10^6$), are not sensitive to ambient magnetic fields, are a fraction of the size and mass, and have comparable resolution and cost.

The energy is calculated from the geometric mean of the $L$ and $R$ gain-corrected, pedestal-subtracted pulse heights from each SiPM, where energy = $\sqrt{(L*R)}$.

We compared quad SiPM data (Fig. 3, *green*) to test data from Fermi LAT calorimeter development (20 mm x 27 mm x 326 mm CsI:Tl crystal, dual PIN photodiode on both ends) and Tetratex diffusive wrap (Fig. 3, *red*). We found that replacing the PIN diodes with SiPMs improves the calorimetry performance for ~10 MeV - 100 MeV pair telescope and enables ~1 MeV - 30 MeV Compton telescope with a modest channel count using macroscopic crystals.

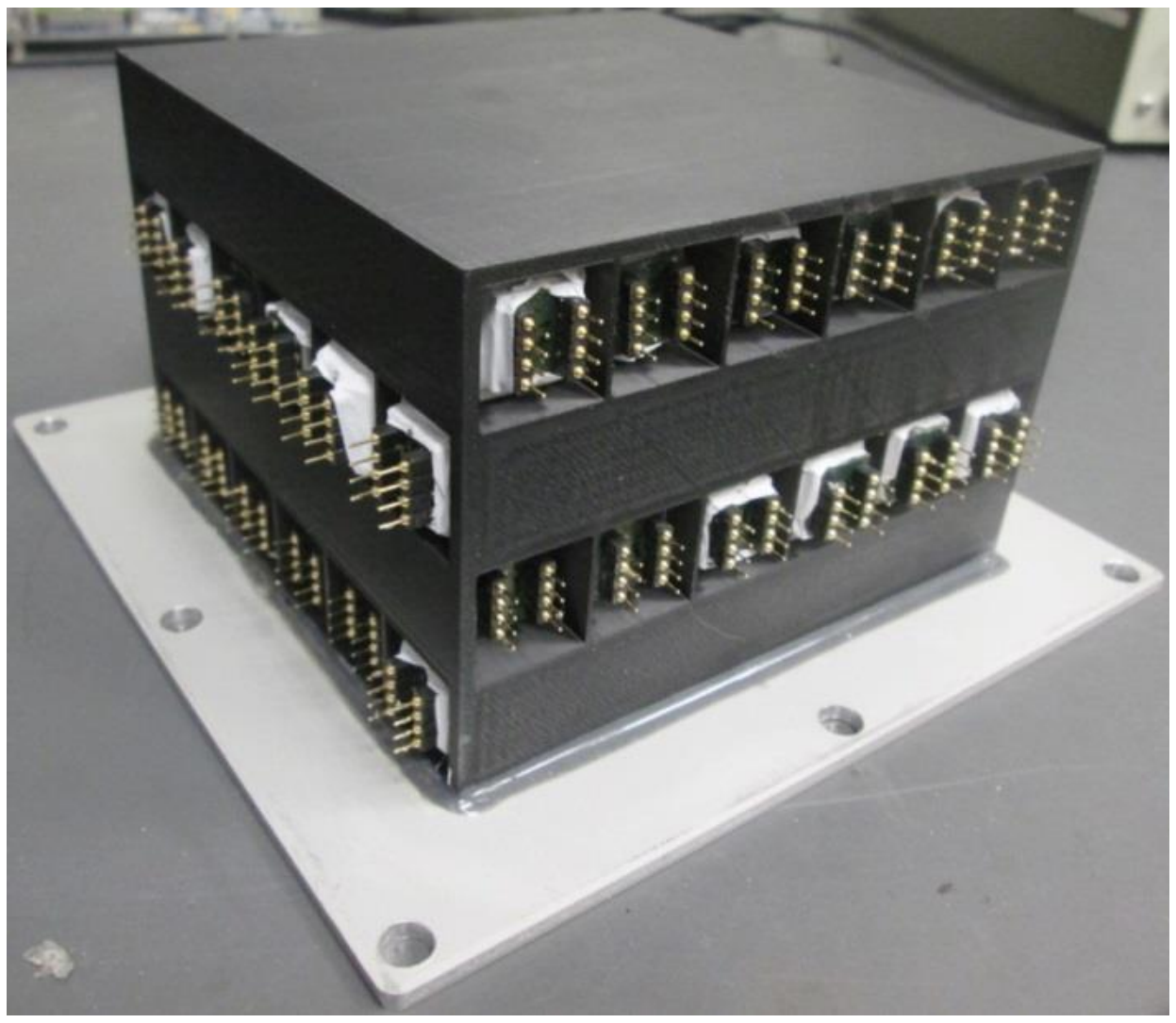

Fig. 4. Fully populated hodoscope consisting of 24-element CsI:Tl CDEs.

## B. *Calorimeter Development*

We have assembled, calibrated and tested three calorimeter prototypes, one of which is intended for future testing with AMEGO Si and CZT prototypes.

The main hodoscope calorimeter, intended for a beam test and a balloon flight, consists of four orthogonally-oriented layers, with each layer containing six 16.7 mm x 16.7 mm x 100 mm crystals, for a total of 24 CsI:Tl crystals (Fig. 4). The crystals were manufactured and polished by St. Gobain crystals. The total thickness of the calorimeter corresponds to 3.7 radiation lengths. Each crystal end is read out by a SensL ArrayJ 2x2 SiPM. The crystals are housed in a 3-*d* printed mechanical structure with a total volume of 118 mm x 118 mm x 71 mm. The hodoscope contains short (100 mm) crystals to achieve comparable cross section with the single-wafer double-sided Si beam test tracker, and will be used for a subsequent beam test and balloon flight.

Additionally, we constructed two single layer arrays of CsI:Tl bars for other testing in the laboratory and at a beam test. The first of which is one layer of six crystals. These crystals consist of the same crystal + SiPM readout described in the previous paragraph. The main purpose of this array is to test the performance of an ASIC specifically designed to handle the large capacitance output of the SiPM array, namely the IDEAS IDE 3380 (SIPHRA) [10]. The IDEAS SIPHRA (Silicon Photomultiplier Readout ASIC) is a 16-channel self-triggering, low-power ASIC intended for the high capacitance of SiPMs coupled to the IDEAS Galao ROIC Development Kit (Fig. 5).

The other single layer array consists of 16.7 mm x 16.7 mm x 430 mm CsI:Tl bars with 12 SensL ArrayJ quad SiPMs in a 3-*d* printed mechanical structure. This bench test calorimeter contains 430 mm (in length) crystals that are consistent with the dimensions of the flight calorimeter concept (see section III).

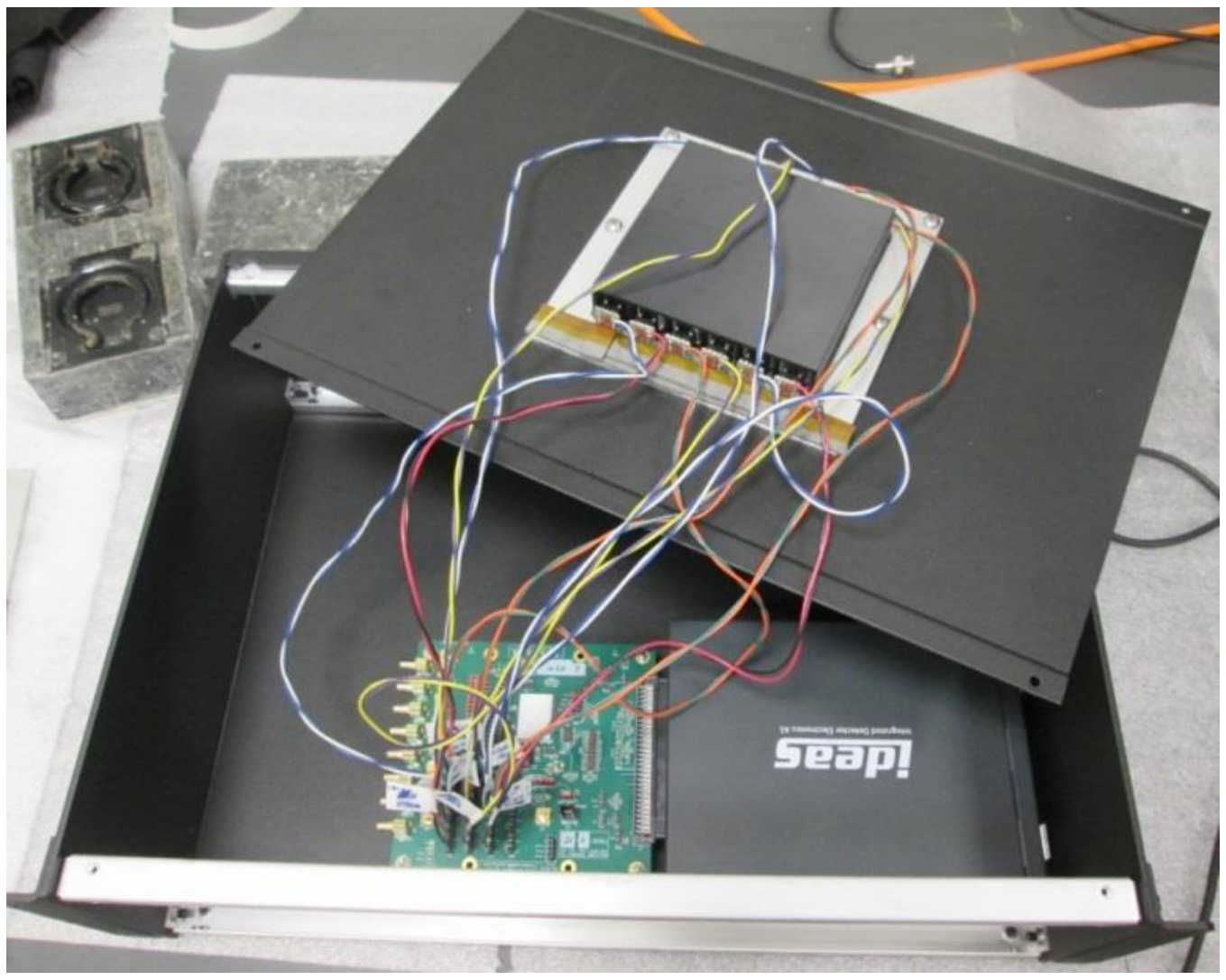

Fig. 5. Fully populated one-layer calorimeter shown with IDEAS SIPHRA + Galao ROIC Development Kit.

## C. *Assembly of Prototypes*

### 1) *Crystal Detector Elements (CDEs)*

A crystal detector element (CDE) consists of the CsI:Tl crystal with optical surfaces treated, one SiPM bonded to each end, and optical closeout wrapping applied (Fig. 6). Each CDE is a testable unit. This section describes the construction process of the CDE is greater detail.

First, we will discuss the optical and mechanical bond between SiPM and CsI:Tl crystal. Several lessons were learned from the development of Fermi LAT, including the large differential coefficient of thermal expansion between CsI:Tl and Si readout will cause failure of the optical bond made with hard optical epoxies, resulting in a reduction of the detected scintillation light by a factor of ~2.

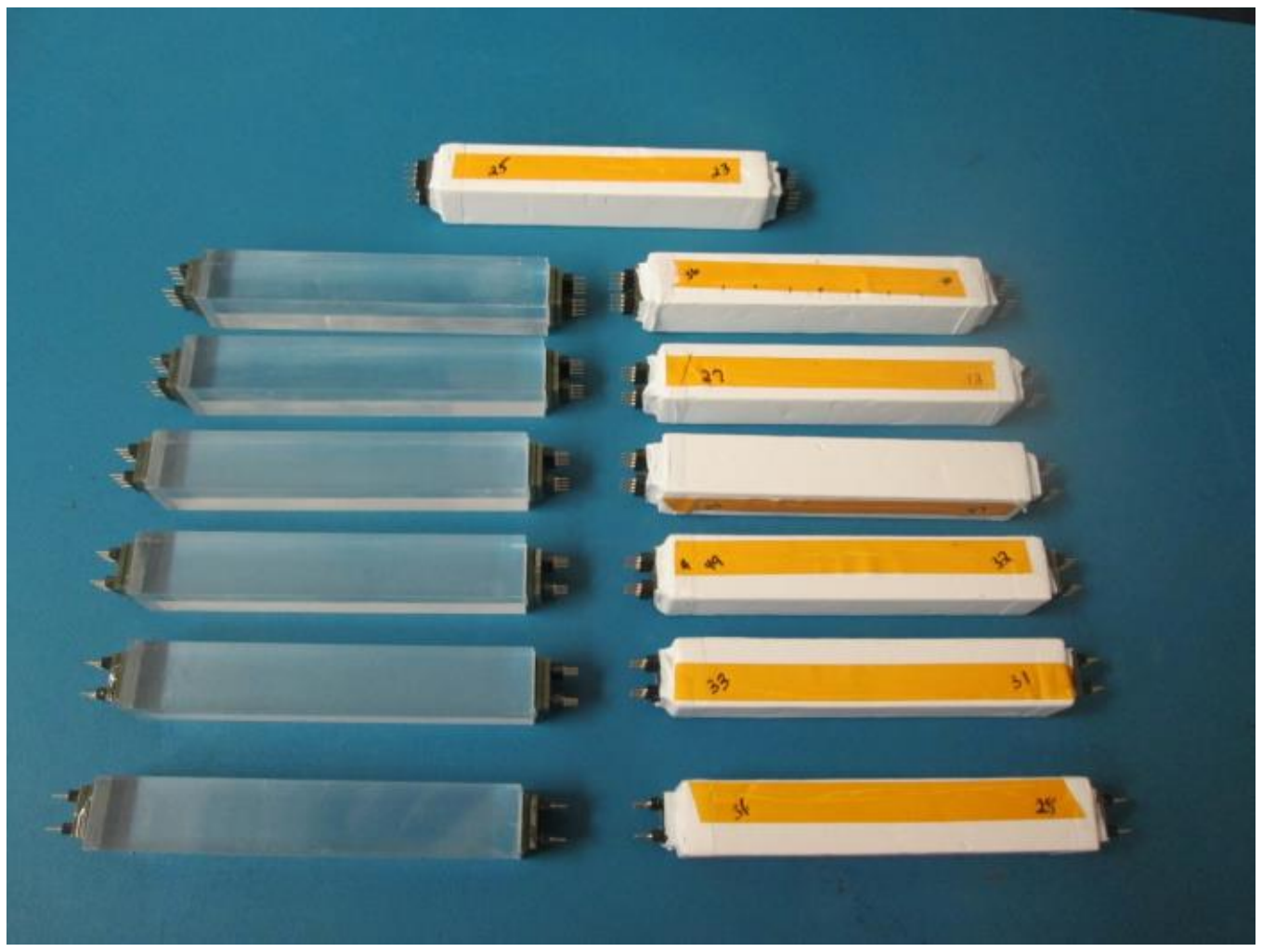

Fig. 6. CsI:Tl crystals with SiPMs bonded to each end, and with Tetratex wrap/PTFE close out window shown to the right.

Laboratory testing revealed that thick (~1 mm) elastomeric bonds (e.g., with Dow Corning DC93-500 silicone encapsulant) will not fail as is currently demonstrated by the LAT calorimeter flight and spare modules in which none of the 3,648 optical bonds have failed.

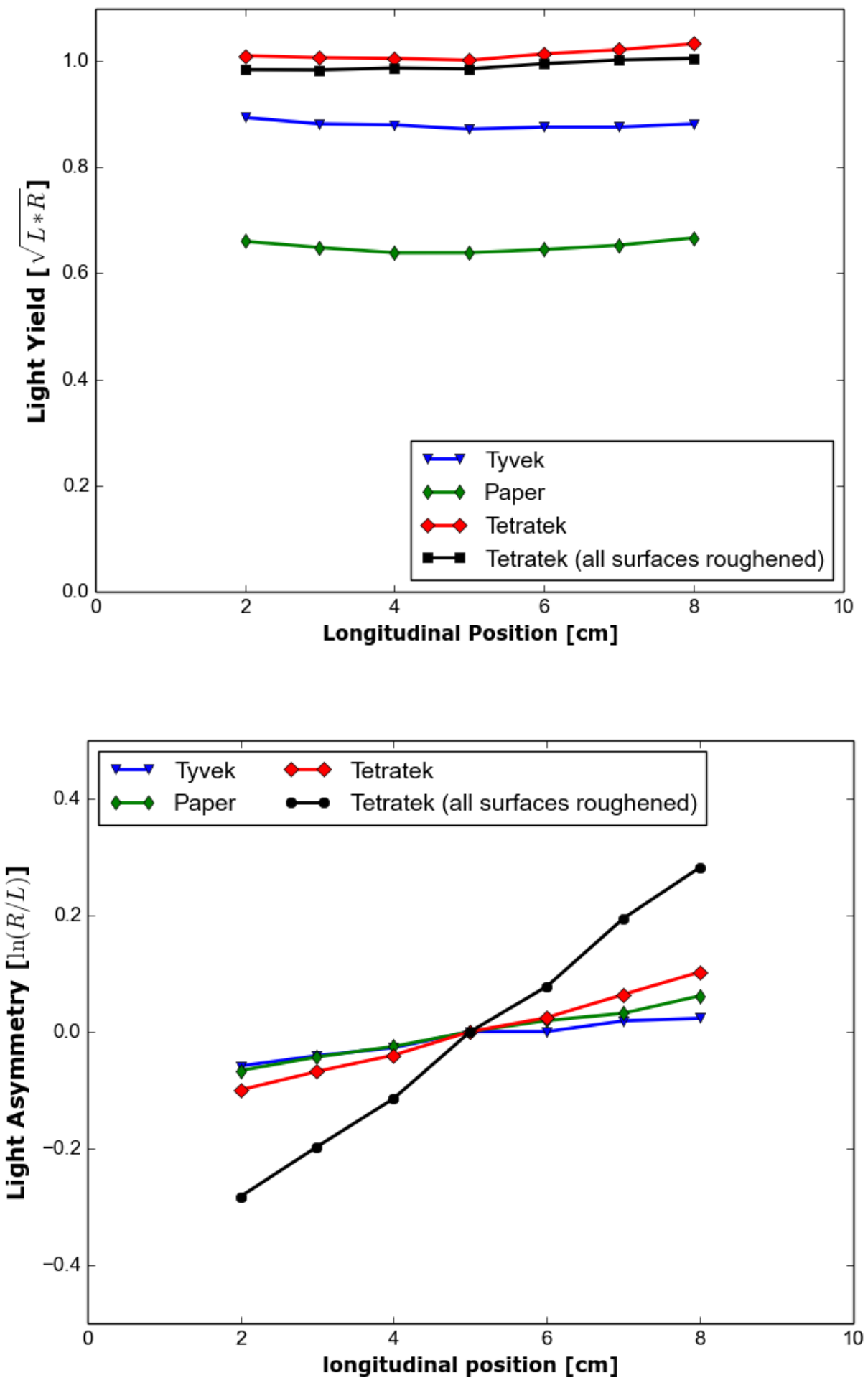

Fig. 7. Results from a CDE with SiPMs bonded on each end of a polished crystal and crystal with roughened surface for varying reflective wraps (white paper, Tyvek, and Tetratex) in terms of light yield (*top*) and light "asymmetry" (*bottom*). The light yield units are arbitrary with the center point of the roughened Tetratex wrap normalized to unity.

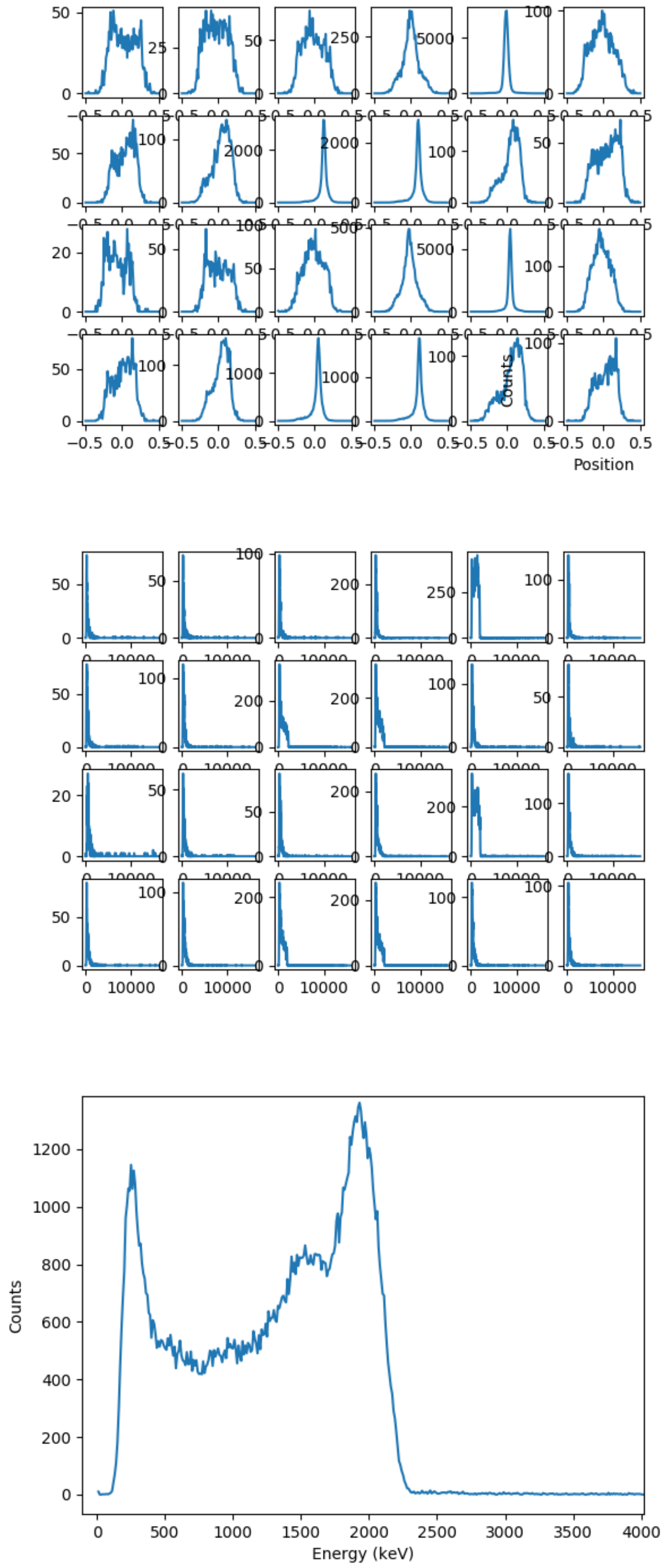

Fig. 8. The reconstructed position (*top*) and energy deposited (*center*), respectively, for each bar in the hodoscope. *Bottom* plot shows the summed total energy deposited in the hodoscope as measured by all 24 CDEs with common calibration coefficient applied to all channels. These data were collected with VME-based Flash ADC digital electronics.

### 2) *Surface Treatment and Optical Closeout*

All sides of each crystal surface that comprise a CDE were treated by roughening with 180-grit sandpaper. Roughening the polished surfaces increases the light attenuation along the length of the crystal, thereby improving position sensitivity for hodoscopic calorimetry while not significantly degrading light output. Several diffusive white wrapping materials were tested for the optical closeout surrounding each crystal; these materials include white paper, Tyvek, and PTFE-based Tetratex. Characterization and performance results are described in the next section.

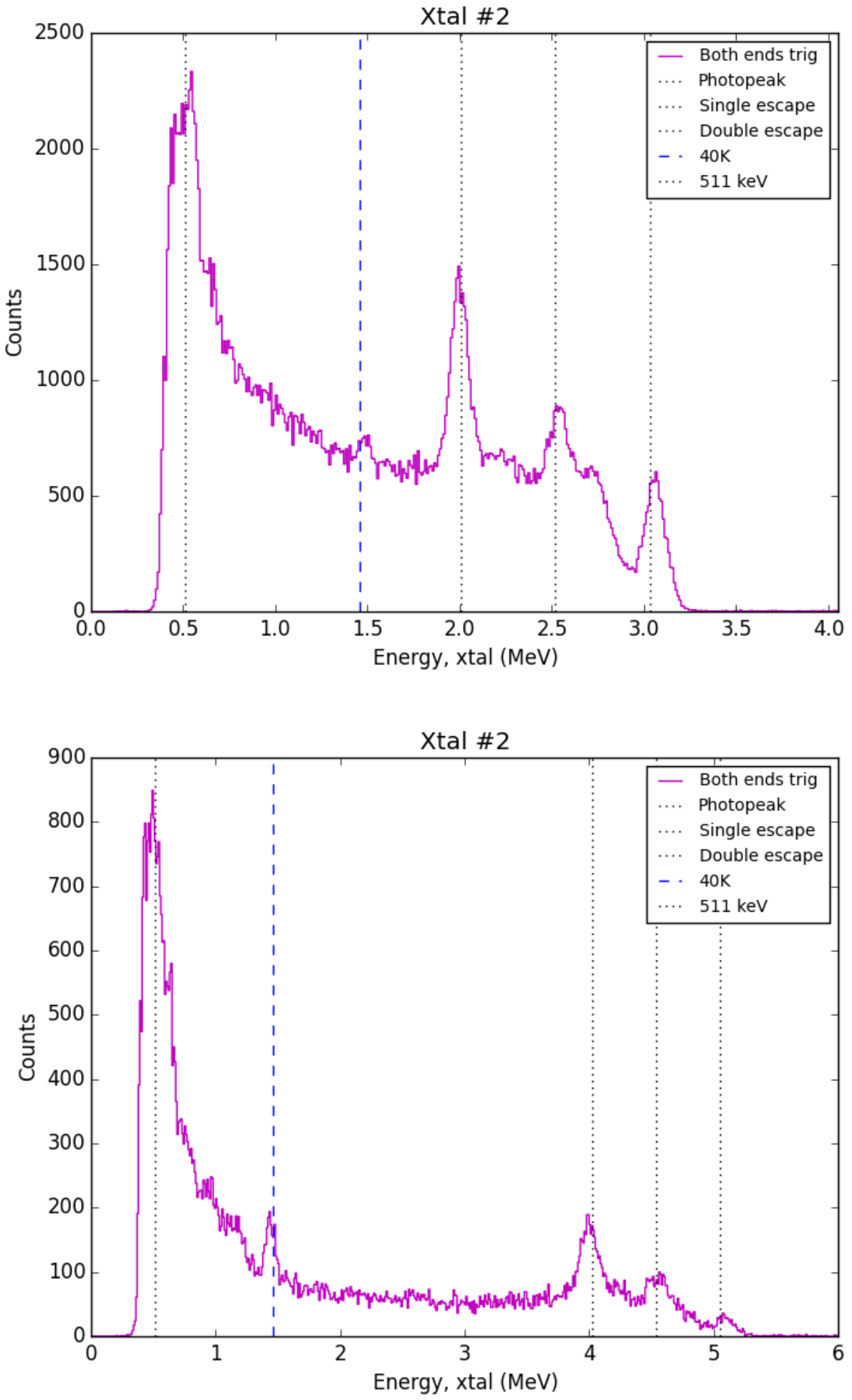

Fig. 9. To test the capabilities of the electronics more suitable for a balloon flight (and subsequent beam test), we used the IDEAS SIPHRA + Galao ROIC Development Kit to read out signals from six CDEs (12 SiPMs total). Shown are the reconstructed 3.03 MeV (*top*) and 5.05 MeV (*bottom*) spectra from one of the six crystals. The RMS resolution at 2 MeV (double escape peak) is 2.36%, or 5.5% FWHM.

### D. Crystal Characterization

#### 1) Light Yield and Asymmetry

Fig. 7, *top* shows that the Tetratex PTFE-based wrap is "whitest" and provides highest light output. The nearly horizontal light asymmetry (blue, red, green curves in Fig. 7, *bottom*) demonstrates the need for roughening the polished crystal surfaces as these three wraps on the polished surface show little taper. When roughened and using the Tetratex for the closeout, the light asymmetry increases dramatically (up to ~60% from one end to the other, compared to ~20% for the same Tetratex closeout on a polished crystal). Roughening all surfaces improves position (RMS) and energy (σ) resolution, 5 mm and 2.8% at 662 keV, respectively. These results were completed and replicated for all 30 100-mm-long CDEs used amongst the two calorimeters.

#### 2) Beam Test

In May 2018, we participated in a beam test at the High Intensity Gamma-ray Sources (HIGS) [11] on the campus of Duke University. The HIGS facility provided mono-energetic γ-ray beams in the energy range from 2.02 MeV – 29.5 MeV. Beyond 29.5 MeV is possible at HIGS but requires a reconfiguration of the beamline mirrors and was not in the scope of the initial beam test. As noted in Section II, we had two calorimeter systems at HIGS – the hodoscope and the one-layer array, each containing the 100-mm-long bars. For the beam test, the 48 channels from the hodoscope were read out with three Struck SIS3316 Flash ADCs [12]. An external bias voltage supply provided the 29.5 V – 5 V above breakdown voltage – needed by each SiPM. The 12 channels from the single layer were readout with the IDEAS SIPHRA + Galao. A separate benchtop power supply provided the bias to the SiPMs read out by the IDEAS system. The beam energies sampled were 2.02 MeV, 3.03 MeV, 5.05 MeV, 10 MeV, 15 MeV, 20 MeV, and 29.5 MeV. Figure 8 shows the reconstructed position (*top*) and energy (*middle*) for each crystal in the hodoscope for the 2.02 MeV beam, requiring coincident hits between two SiPMs on the same crystal. The reconstructed position and energy display which crystal was irradiated by the beam (spot size = 8 mm), demonstrating a sharp peak (in position) and photopeak + single and double escape peaks (in energy), while crystals that were not directly in the beam display measurements that are consistent with background. Figure 8 (*bottom*) shows the summation of all crystals in the hodoscope for the 2.02 MeV beam.

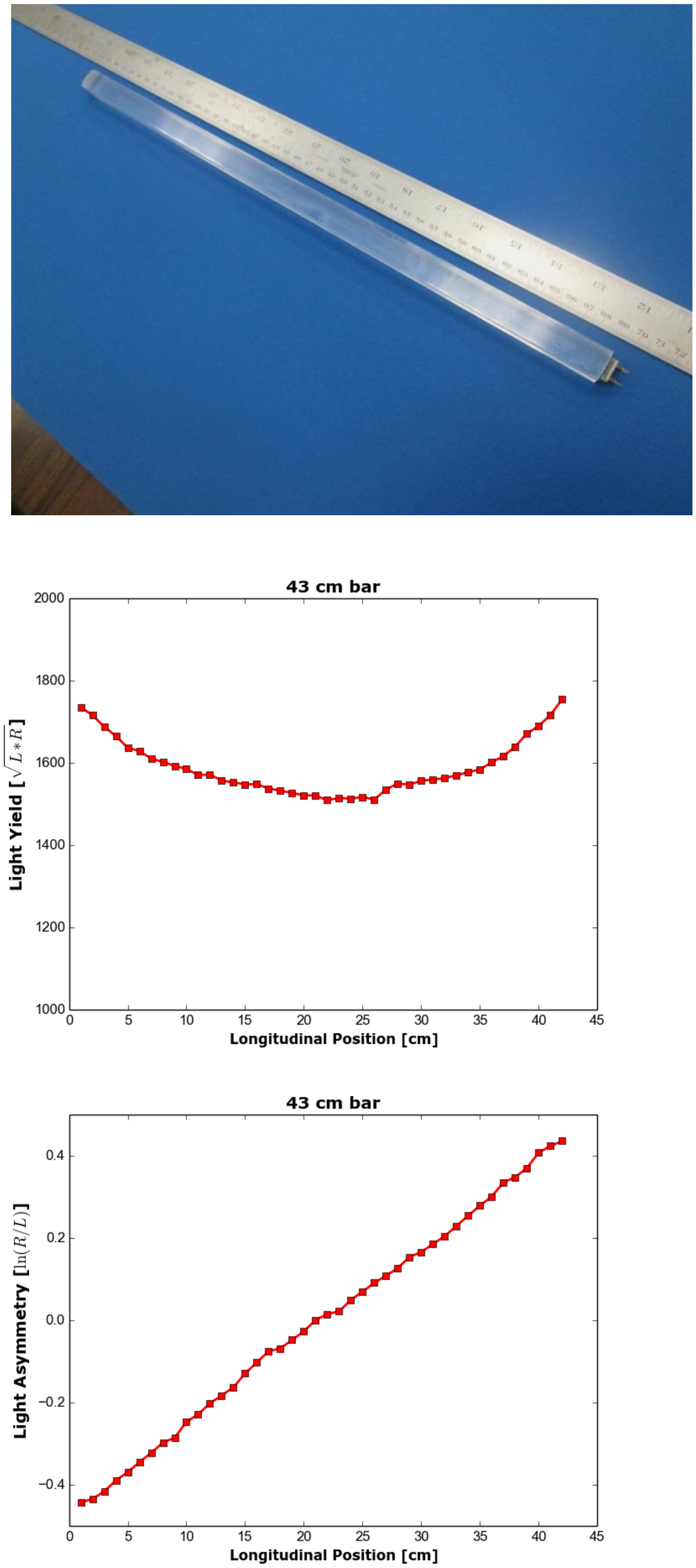

Fig. 10. Polished 16.7 mm x 16.7 mm x 430 mm CsI:Tl bar with SensL ArrayJ quad SiPMs bonded to each end (*top*). The light yield (*middle*) and light "asymmetry" (*bottom*) plotted as a function of position.

Figure 9 shows the results at 3.03 MeV (*top*) and 5.05 MeV (*bottom*) measured by one crystal read out by two channels of the IDEAS SIPHRA + Galao. For each energy the photopeak was clearly measured, along with the single and double escape peaks and other background lines. The gain and non-linearity were derived from the combination of the data acquired at these beam energies. The RMS resolution at 2 MeV (from the double escape peak) is 2.36% (σ) or 5.5% FWHM.

### 3) *Crystal for AMEGO Probe-Class Mission*

As previously mentioned, the length of the CsI:Tl bar for the beam test and balloon flight was mainly driven by the size of the DSSD tracker and CZT subsystems (~100 cm$^2$). Longer CsI:Tl crystals – with a length on the order of 50 cm – would be required for the AMEGO Probe-class mission.

At the end of 2017, the longest CsI:Tl – with 16.7 mm x 16.7 mm cross section – that we could procure (from ScintiTech crystals) was 430 mm in length (Fig. 10, *top*). We have begun the initial tests using this crystal. Thus far we have left the surface polished, bonded quad SiPMs to each end and wrapped the bar with Tetratex in an effort to understand the response in terms of light yield and asymmetry. Fig. 10 (*middle* and *bottom*) show the mapping of the bar in terms 10-mm-step sizes using a collimated $^{137}$Cs source. These results show the irregularities in the response at this scale. Further testing with different wrapping and surface treatments will be completed in 2019.

## IV. Future Work

As was demonstrated in this paper, the beam test / balloon flight calorimeter has been built, tested and initially calibrated in the laboratory and at a first beam test. The main area of work we are now focused on will be implementing a 64-channel version of the IDEAS SIPHRA, known as the IDEAS ROSSPAD [13]. The ROSSPAD is essentially four SIPHRA on a common board and with the similar functionality as the single SIPHRA. The first half of 2019 will be spent working on implementing the ROSSPAD DAQ with the CsI:Tl calorimeter and integrating it with the DSSD tracker and CZT subsystems. The beam test with all three subsystems is planned for late 2019 at the HIGS facility, followed by a balloon flight at a location TBD in late 2020.

## Acknowledgment

We are grateful to the accelerator physics group led by Prof. Y.K. Wu at High Intensity Gamma-ray Source (HIGS), Triangle Universities Nuclear Laboratory for developing and producing an excellent γ-ray beam for the tests. The HIGS facility is supported in part by U.S. Department of Energy grant DE-FG02-97ER41033. Additionally, we are grateful for the work done by Naval Research Enterprise Internship Program (NREIP) summer intern student, Mr. Andrew Maris.